\documentclass[11pt]{article}
\usepackage{DGfest,epsfig}
\usepackage{times}
\usepackage{amsmath}
\usepackage{amssymb}
\usepackage{graphicx}
\usepackage{tabularx}
\def\be{\begin{equation}}
\def\ee{\end{equation}}
\def\bea{\begin{eqnarray}}
\def\eea{\end{eqnarray}}

\begin{document}
\baselineskip 11.5pt
\title{PROBING THE QCD VACUUM USING EXTERNAL FIELDS}

\author{ Paolo Cea$^{1,2}$  and Leonardo Cosmai$^{1}$}
\address{$^1$INFN - Sezione di Bari, via Amendola 173, 70126 Bari, Italy\\
$^2$Dipartimento di Fisica, Universit\`a di Bari,  via Amendola 173, 70126 Bari, Italy}
%\\

\maketitle\abstracts{
The QCD vacuum can be studied using external fields. We report here results respectively obtained
probing the lattice QCD vacuum by means of an abelian monopole field and of an abelian chromomagnetic field.}

%%%%%%%%%%%%%%%%%%%%%%%%%%%%%%%%%%%%%%%%%%%%%%%%%%%%%%%%%%%%%%%%%%%%%%%%%%%%%%%%%%%%%%%%%%%%%%%%%%%%%%%%%%%%%%%%%%%%%%%%%%%%%%%%%%%%%
\section{Introduction}
%%%%%%%%%%%%%%%%%%%%%%%%%%%%%%%%%%%%%%%%%%%%%%%%%%%%%%%%%%%%%%%%%%%%%%%%%%%%%%%%%%%%%%%%%%%%%%%%%%%%%%%%%%%%%%%%%%%%%%%%%%%%%%%%%%%%%
Understanding color confinement is one of the long standing problems
of the Standard Model of Particle Physics.
Notwithstanding many theoretical efforts a totally convincing explanation of the
confinement phenomenon is still lacking (we refer the reader to recent
reviews on confinement~\cite{DiGiacomo:2005yq,Ripka:2003vv,Greensite:2003bk,Haymaker:1998cw}) and a full
understanding of the QCD vacuum dynamics  is not yet at our disposal.
Nevertheless lattice computations can shed light on many aspects of the QCD vacuum
and possibly help in identifying the mechanism of the color
confinement:
that has surely been a leading interest in Adriano's research activity.
In this paper we review some results we have collected in the last few
years in studying the QCD vacuum by means of external fields.
The plan of the paper is the following. In section~\ref{extfields} we briefly recall our gauge invariant
method to put external fields on the lattice. In section~\ref{abmonopoles} we consider the QCD vacuum
under the influence of an Abelian monopole external field and we study the phase transition, even with
two dynamical flavors,  using
the free energy in presence of the monopole field.
In section~\ref{abchromomagfield} we probe the vacuum structure by means of a
constant abelian chromomagnetic field, both at zero and finite temperature.
Finally in section~\ref{conclusions} we present our conclusions.

%%%%%%%%%%%%%%%%%%%%%%%%%%%%%%%%%%%%%%%%%%%%%%%%%%%%%%%%%%%%%%%%%%%%%%%%%%%%%%%%%%%%%%%%%%%%%%%%%%%%%%%%%%%%%%%%%%%%%%%%%%%%%%%%%%%%%
\section{External Fields on the Lattice}
\label{extfields}
%%%%%%%%%%%%%%%%%%%%%%%%%%%%%%%%%%%%%%%%%%%%%%%%%%%%%%%%%%%%%%%%%%%%%%%%%%%%%%%%%%%%%%%%%%%%%%%%%%%%%%%%%%%%%%%%%%%%%%%%%%%%%%%%%%%%%

%%%%%%%%%%%%%%%%%%%%%%%%%%%%%%%%%%%%%%%%%%%%%%%%%%%%%%%%%%%%%%%%%%%%%%%%%%%%%%%%%%%%%%%%%%%%%%
\subsection{Pure gauge theories at zero temperature. The lattice effective action}
%%%%%%%%%%%%%%%%%%%%%%%%%%%%%%%%%%%%%%%%%%%%%%%%%%%%%%%%%%%%%%%%%%%%%%%%%%%%%%%%%%%%%%%%%%%%%%

In previous papers~\cite{Cea:1997ff,Cea:1999gn} we introduced a lattice
effective action $\Gamma[\vec{A}^{\text{ext}}]$ for an external background
field $\vec{A}^{\text{ext}}$, which is gauge invariant against static gauge
transformations of the background field
\be
\label{Gamma}
\Gamma[\vec{A}^{\text{ext}}] = -\frac{1}{L_t} \ln
\left\{
\frac{{\mathcal{Z}}[\vec{A}^{\text{ext}}]}{{\mathcal{Z}}[0]}
\right\}
\ee
where $L_t$ is the lattice size in time direction and
$\vec{A}^{\text{ext}}(\vec{x})$ is the continuum gauge potential of the
external static background field.  ${\mathcal{Z}}[\vec{A}^{\text{ext}}]$ is the
lattice partition functional
\be \label{Zetalatt}
{\mathcal{Z}}[\vec{A}^{\text{ext}}] =
\int_{U_k(\vec{x},x_t=0)=U_k^{\text{ext}}(\vec{x})} {\mathcal{D}}U \; e^{-S_W}
\,, \ee
with $S_W$ the standard pure gauge Wilson action.

The functional integration is performed over the lattice links, but constraining
the spatial links belonging to a given time slice (say $x_t=0$) to be
\be
\label{coldwall}
U_k(\vec{x},x_t=0) = U^{\text{ext}}_k(\vec{x})
\,,\,\,\,\,\, (k=1,2,3) \,\,,
\ee
 $U^{\text{ext}}_k(\vec{x})$ being the lattice version  of the external continuum
gauge potential $\vec{A}^{\text{ext}}(x)=\vec{A}^{\text{ext}}_a(x) \lambda_a/2$.
Note that the temporal links are not constrained.

In the case of a static background field which does not vanish at infinity we
must also impose that, for each time slice $x_t \ne 0$, spatial links exiting
from sites belonging to the spatial boundaries  are fixed according to
eq.~(\ref{coldwall}). In the continuum this last condition amounts to the
requirement that fluctuations over the background field vanish at infinity.

The partition function defined in eq.~(\ref{Zetalatt}) is also known as lattice
Schr\"odinger functional~\cite{Luscher:1992an} and in the
continuum corresponds to the Feynman kernel~\cite{Rossi:1980jf}. Note that, at
variance with the usual formulation of the lattice Schr\"odinger
functional~\cite{Luscher:1992an} where a lattice cylindrical
geometry is adopted, our lattice has an hypertoroidal geometry so that $S_W$ in
eq.~(\ref{Zetalatt}) is allowed to be the standard Wilson action.

The lattice effective action $\Gamma[\vec{A}^{\text{ext}}]$ corresponds to the vacuum
energy, $E_0[\vec{A}^{\text{ext}}]$,
in presence of the background field with respect to the vacuum energy, $E_0[0]$,
with  $\vec{A}^{\text{ext}}=0$.

%%%%%%%%%%%%%%%%%%%%%%%%%%%%%%%%%%%%%%%%%%%%%%%%%%%%%%%%%%%%%%%%%%%%%%%%%%%%%%%%%%%%%%%%%%%%%%
\subsection{Pure gauge theories at finite temperature.  The thermal partition functional}
%%%%%%%%%%%%%%%%%%%%%%%%%%%%%%%%%%%%%%%%%%%%%%%%%%%%%%%%%%%%%%%%%%%%%%%%%%%%%%%%%%%%%%%%%%%%%%

If we now consider the gauge theory at finite temperature $T=1/(a L_t)$
in presence of an external background field, the relevant quantity is
the free energy functional defined as
\be
\label{freeenergy}
{\mathcal{F}}[\vec{A}^{\text{ext}}] = -\frac{1}{L_t} \ln
\left\{
\frac{{\mathcal{Z_T}}[\vec{A}^{\text{ext}}]}{{\mathcal{Z_T}}[0]}
\right\} \; .
\ee
${\mathcal{Z_T}}[\vec{A}^{\text{ext}}]$ is the thermal partition
functional~\cite{Gross:1981br}
in presence of the background field $\vec{A}^{\text{ext}}$, and is defined as
\be
\label{ZetaTnew}
\mathcal{Z}_T \left[ \vec{A}^{\text{ext}} \right]
= \int_{U_k(\vec{x},L_t)=U_k(\vec{x},0)=U^{\text{ext}}_k(\vec{x})}
\mathcal{D}U \, e^{-S_W}   \,.
\ee
In eq.~(\ref{ZetaTnew}), as in eq.~(\ref{Zetalatt}), the spatial links
belonging to the time slice $x_t=0$ are constrained to the value of the
external background field, the temporal links are not constrained.
The free energy functional eq.~(\ref{freeenergy})
corresponds to the free energy, $F[\vec{A}^{\text{ext}}]$, in presence of the
external background field evaluated with respect to the free energy, $F[0]$,
with $\vec{A}^{\text{ext}}=0$. When the physical temperature is sent to zero
the free energy  functional reduces to the vacuum energy functional
eq.~(\ref{Gamma}).

%%%%%%%%%%%%%%%%%%%%%%%%%%%%%%%%%%%%%%%%%%%%%%%%%%%%%%%%%%%%%%%%%%%%%%%%%%%%%%%%%%%%%%%%%%%%%%
\subsection{Including dynamical fermions}
%%%%%%%%%%%%%%%%%%%%%%%%%%%%%%%%%%%%%%%%%%%%%%%%%%%%%%%%%%%%%%%%%%%%%%%%%%%%%%%%%%%%%%%%%%%%%%

When including dynamical fermions, the thermal partition functional
in presence of a static external background gauge field, Eq.~(\ref{ZetaTnew}),
becomes:
\begin{eqnarray}
\label{ZetaTfermions}
\mathcal{Z}_T \left[ \vec{A}^{\text{ext}} \right]  &=&
\int_{U_k(L_t,\vec{x})=U_k(0,\vec{x})=U^{\text{ext}}_k(\vec{x})}
\mathcal{D}U \,  {\mathcal{D}} \psi  \, {\mathcal{D}} \bar{\psi} e^{-(S_W+S_F)}
\nonumber \\
&=&  \int_{U_k(L_t,\vec{x})=U_k(0,\vec{x})=U^{\text{ext}}_k(\vec{x})}
\mathcal{D}U e^{-S_W} \, \det M \,,
\end{eqnarray}
where $S_W$ is the Wilson action, $S_F$ is the fermionic action and $M$ is
the fermionic matrix.
Notice that the fermionic fields are not constrained and
the integration constraint is only relative to the gauge fields:
this leads, as in the usual QCD partition function, to the appearance of
the gauge invariant fermionic determinant after integration on the
fermionic fields.
As usual we impose on fermionic fields
periodic boundary conditions in the spatial directions and
antiperiodic boundary conditions in the temporal direction.

%%%%%%%%%%%%%%%%%%%%%%%%%%%%%%%%%%%%%%%%%%%%%%%%%%%%%%%%%%%%%%%%%%%%%%%%%%%%%%%%%%%%%%%%%%%%%%%%%%%%%%%%%%%%%%%%%%%%%%%%%%%%%%%%%%%%%
\section{Abelian Monopoles}
%%%%%%%%%%%%%%%%%%%%%%%%%%%%%%%%%%%%%%%%%%%%%%%%%%%%%%%%%%%%%%%%%%%%%%%%%%%%%%%%%%%%%%%%%%%%%%%%%%%%%%%%%%%%%%%%%%%%%%%%%%%%%%%%%%%%%
\label{abmonopoles}

A mechanism for color confinement based on dual superconductivity of the
QCD vacuum by abelian monopole condensation has been proposed
a long time ago~\cite{tHooft:1976eps,Mandelstam:1974pi}.
A disorder parameter which is related to abelian monopole condensation
in the dual superconductivity picture of confinement has been
developed by the Pisa group and consists in the vacuum expectation value of a
magnetically charged operator, $\langle \mu \rangle$.

Our proposal is to detect abelian monopole condensation  by looking at the
free energy~\cite{Cea:2000zr,Cea:2001an}
in presence of an abelian monopole background field.

Since the free energy is related to the
vacuum dynamics and not, like the trace of the Polyakov loop, to a symmetry of the
gauge action which is washed out by the presence of dynamical fermions,
we feel that it could be used
to detect the finite temperature phase transition also in full QCD.

%%%%%%%%%%%%%%%%%%%%%%%%%%%%%%%%%%%%%%%%%%%%%%%%%%%%%%%%%%%%%%%%%%%%%%%%%%%%%%%%%%%%%%%%%%%%%%
\subsection{Abelian monopole condensation in pure gauge theories}
%%%%%%%%%%%%%%%%%%%%%%%%%%%%%%%%%%%%%%%%%%%%%%%%%%%%%%%%%%%%%%%%%%%%%%%%%%%%%%%%%%%%%%%%%%%%%%

For SU(3) gauge theory the maximal abelian group is
U(1)$\times$U(1), therefore we may introduce two independent types
of abelian monopoles using respectively the Gell-Mann matrices
$\lambda_3$ and $\lambda_8$ or their linear combinations.

In the following we shall consider the abelian monopole field related to
the $\lambda_3$ diagonal generator.
In the continuum the abelian monopole field is given by
\begin{equation}
\label{monop3su2}
g \vec{b}^a({\vec{x}}) = \delta^{a,3} \frac{n_{\mathrm{mon}}}{2}
\frac{ \vec{x} \times \vec{n}}{|\vec{x}|(|\vec{x}| -
\vec{x}\cdot\vec{n})} \,,
\end{equation}
where $\vec{n}$ is the direction of the Dirac string and,
according to the Dirac quantization condition, $n_{\text{mon}}$ is
an integer. The spatial lattice links are constrained (see Eq.~(\ref{coldwall}))
to the values corresponding to the continuum abelian monopole field Eq.~(\ref{monop3su2}).

% FIGURE 1
\begin{figure}[ht]
\includegraphics[width=0.85\textwidth,clip]{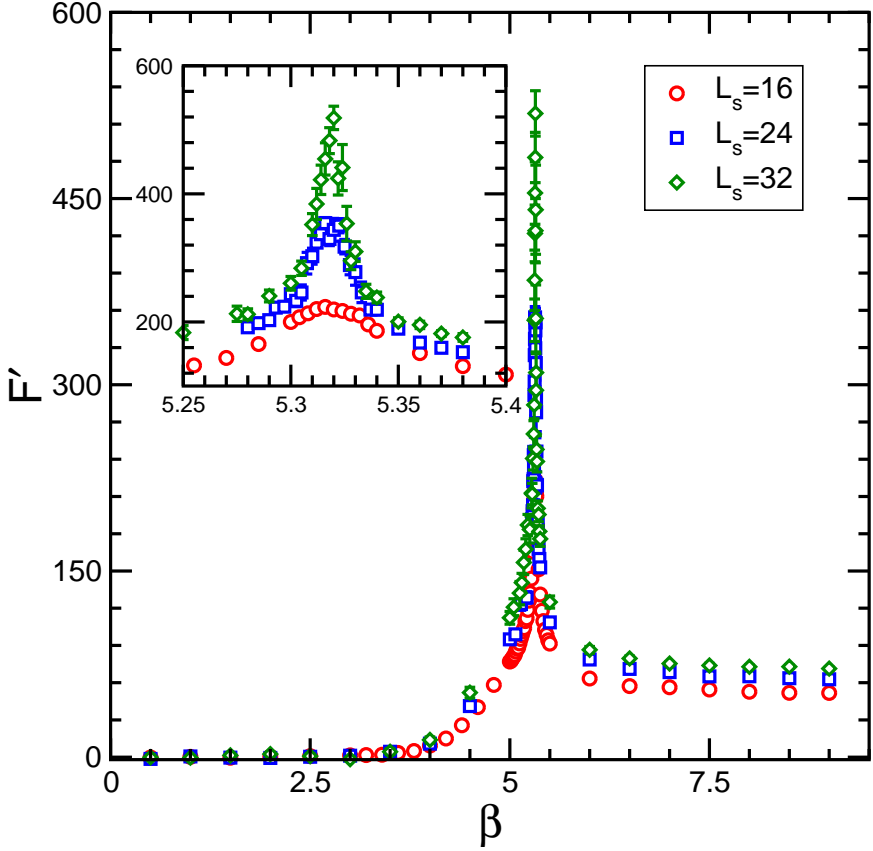}
\caption{
$F^{\prime}$ vs.  $\beta$  ($n_{\text{mon}}=10$)
for $L_s=16,24,32$ and $L_t=4$.}
\label{Fig1}
\end{figure}

% FIGURE 2
\begin{figure}[ht]
\includegraphics[width=0.85\textwidth,clip]{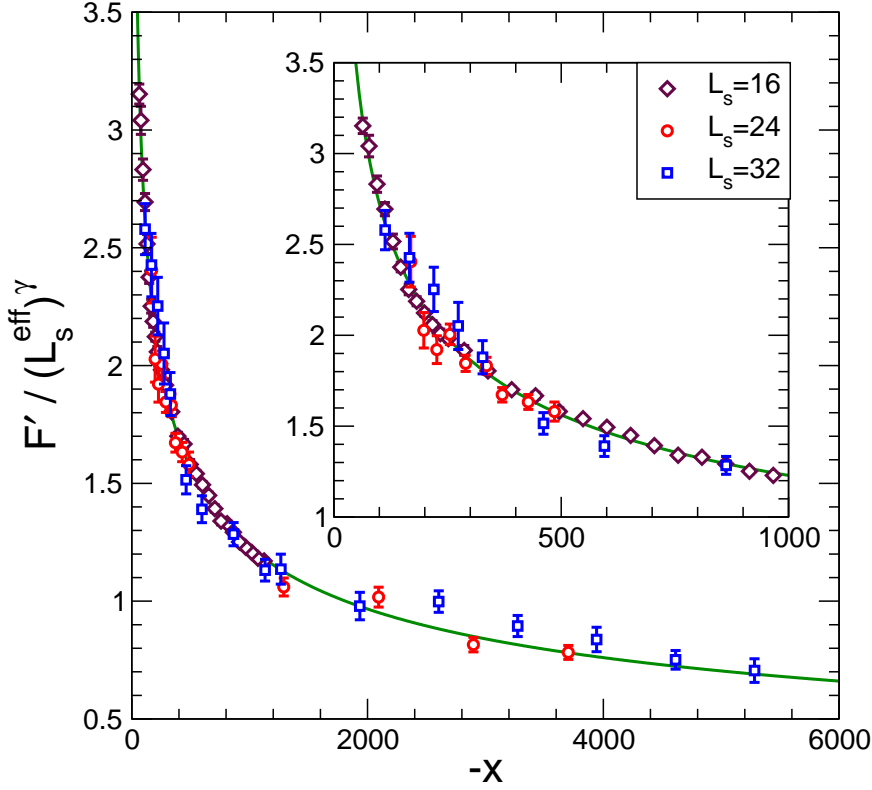}
\caption{The derivative of the monopole free energy with respect to the gauge coupling
$\beta$ rescaled with  $(L_s^{\text{eff}})^{\gamma}$ versus the
scaling variable $x$.
}
\label{Fig2}
\end{figure}

The monopole background field is introduced
by constraining  the spatial links exiting from the sites
at the boundary of the time slice $x_t=0$ (see Eq.~(\ref{coldwall}). For what concern spatial links exiting from sites
at the boundary of other time slices ($x_t \ne 0$), since the monopole field vanishes at
infinity,  we consider two possibilities.
In the first one we constrain these links according to Eq.~(\ref{coldwall}) (in the
following we refer to this possibility as ``fixed boundary conditions'').
In the second one we do not impose the constraint Eq.~(\ref{coldwall}) on the above mentioned links
(this possibility will be referred as ``periodic boundary conditions'').

We simulate pure SU(3) lattice gauge theory with Wilson action. The lattice geometry is
hypertoroidal.
In Fig.~\ref{Fig1} we report our numerical results for
the derivative of the free energy $F^\prime$.
The data display a sharp peak that increases by increasing the lattice spatial
volume.
By inspecting Figure~\ref{Fig1} it is evident that $F^\prime(\beta)=0$
in a finite range of $\beta$, starting from $\beta=0$ and below the critical coupling,
signaled by a peak in $F^\prime(\beta)$.
Therefore $F(\beta)=0$ in a finite range below the critical coupling,  above which
the gauge system
gets deconfined. The vanishing of the free energy implies abelian monopole condensation.
On the other hand, $F^\prime(\beta)$ becomes different from zero and increases with the lattice
spatial volume near the critical coupling, as expected in presence of a phase transition.
Moreover in the weak coupling regime $F^\prime(\beta)$ stays
constant and almost independent from the spatial lattice volume.
This corresponds to the
classical monopole energy which depends linearly on $\beta$. It is clear that
in the deconfined phase
it costs a finite amount of energy to create a monopole, and for,  there is not
abelian monopole condensation.

A finite size scaling analysis can be performed and it can be shown that
the scaling parameters are compatible with a first order phase transition.
As a first step we determine the value of the critical coupling $\beta_c(L_s^{\text{eff}})$,
where $L_s^{\text{eff}}$ is the effective spatial size ($L_s^{\text{eff}}=L_s-2$ in the case
of fixed boundary conditions).
We fitted our lattice data with the scaling law
\begin{equation}
\label{scalinglaw}
F^{\prime}(\beta,L_s^{\text{eff}}) = \frac{a_1 (L_s^{\text{eff}})^{\gamma}}{\left|  (L_s^{\text{eff}})^{1/\nu} (\beta - \beta_c) - d_1 \right|^\alpha} \,.
\end{equation}
\begin{table}
\begin{tabularx}{0.85\textwidth}{|XXXXXX|}
\hline
\hline
\multicolumn{6}{|c|}{spatial ``fixed boundary conditions'' } \\ \hline
$a_1$       & $\gamma$   &  $\beta_c$   &    $\nu$   & $d_1$    & $\alpha$   \\ \hline
$12.199$    & $1.247$    &  $5.3251$    & $0.335$    & $0.6$    & $0.351$    \\
$\pm3.9004$ & $\pm0.089$ &  $\pm0.0110$ & $\pm0.026$ & constant & $\pm0.035$ \\ \hline \hline
\multicolumn{6}{|c|}{spatial ``periodic boundary conditions''} \\ \hline
$a_1$       & $\gamma$   &  $\beta_c$   &    $\nu$   & $d_1$    & $\alpha$   \\ \hline
$13.461$    & $1.510$    &  $5.3222$    & $0.340$    & $0.6$    & $0.347$    \\
$\pm1.337$ & $\pm0.555$ &  $\pm0.0013$ & $\pm0.020$ & constant & $\pm0.009$  \\ \hline \hline
\end{tabularx}
\caption{The values of the parameters obtained by fitting Eq.~(\ref{scalinglaw}) to the
data for the derivative of the monopole free energy
on lattices with spatial volumes $16^3$, $24^3$, and $32^3$ and spatial ``fixed'' or ``periodic''
boundary conditions respectively.}
\label{Table2}
\end{table}
The output of the fits are reported in Table~\ref{Table2}. The parameter
$d_1$ has been fixed at the value obtained with the fit Eq.~(\ref{scalinglaw}).
We see clearly that $\alpha/\nu$ agrees with $\gamma$ within statistical errors,
as expected if we want a sensible result in the thermodynamical limit $L_s^{\text{eff}} \to \infty$
(see Eq.~(\ref{scalinglaw})).

In Fig.~\ref{Fig2} we plot $F^{\prime}(\beta,L_s^{\text{eff}})$ rescaled with the factor
$1/(L_s^{\text{eff}})^{\gamma}$
versus the scaling variable $x$, respectively
for spatial ``fixed boundary conditions'' and for spatial ``periodic boundary conditions''.
The full line is
\begin{equation}
\label{scalingcurve}
\frac{F^{\prime}(\beta,L_s^{\text{eff}})}{(L_s^{\text{eff}})^{\gamma}} =
\frac{a_1}{\left|  (L_s^{\text{eff}})^{1/\nu} (\beta - \beta_c) - d_1 \right|^\alpha} \,.
\end{equation}
We find that the scaling relation holds quite well for a very large range of $x$.
The quality of the scaling can be inferred looking at Fig.~\ref{Fig2}.

It is interesting to comment on the behavior of $\exp(-F(\beta)/T)$, which is the
analogous of disorder parameter developed by the Pisa
group~\cite{DiGiacomo:1999fa,DiGiacomo:1999fb,Carmona:2001ja}, in the
thermodynamical limit implied by Eq.~(\ref{scalingcurve}). Indeed we have that
\begin{equation}
\label{Fdibeta}
\exp\left(-\frac{F(\beta)}{T}\right) =
\exp \left( -\frac{1}{T} \int_{\beta_0}^\beta  F^{\prime}(\beta^\prime)\, d \beta^\prime  \right) \quad,
\quad \beta_0 < \beta < \beta_c  \,,
\end{equation}
while we already know that $\exp(-F(\beta)/T)=1$for $\beta < \beta_0$ irrespective of the
lattice size (see Fig.~(\ref{Fig2})).
From Eq.~(\ref{scalinglaw}) we get
\begin{equation}
\label{Fdibetaint}
\exp\left(-\frac{F(\beta)}{T}\right) =  \exp\left[ - \frac{1}{T} \,\frac{a_1}{1-\alpha}
\left( |\beta_0-\beta_c|^{1-\alpha}  - |\beta-\beta_c|^{1-\alpha}  \right)  \right]\,, \quad
\beta_0 < \beta < \beta_c  \,.
\end{equation}
So that $F(\beta)$ decreases when $\beta \to \beta_c$ if $0 < \alpha < 1$ tending to a finite value
at $\beta = \beta_c$. On the other hand, for $\alpha=1$ it is easy
to see that $\exp(-F(\beta)/T)$
decreases to zero as a power of $(\beta_c-\beta)$. Thus we conclude that for $\alpha < 1$
we have a discontinuous jump of $\exp(-F(\beta)/T)$ at $\beta_c$
and the strength of the discontinuity weakens when $\alpha \to 1$.
However, it must be stressed that the discontinuous jump of $\exp(-F(\beta)/T)$
is exceedingly small so that $\exp(-F(\beta)/T)$ decreases almost continuously
toward zero when $\beta \to \beta_c$.

%%%%%%%%%%%%%%%%%%%%%%%%%%%%%%%%%%%%%%%%%%%%%%%%%%%%%%%%%%%%%%%%%%%%%%%%%%%%%%%%%%%%%%%%%%%%%%
\subsection{The deconfinement transition in QCD with two dynamical flavors}
%%%%%%%%%%%%%%%%%%%%%%%%%%%%%%%%%%%%%%%%%%%%%%%%%%%%%%%%%%%%%%%%%%%%%%%%%%%%%%%%%%%%%%%%%%%%%%

We compute the derivative of the monopole background field free energy
with respect to the gauge coupling
where now the expectation value is evaluated with the full QCD action.
Our main goal is to try to use our data for $F^{\prime}(\beta)$ on different spatial
volumes and different bare quark masses to infer the critical behavior of two flavors
full QCD  near the deconfining transition (for details on numerical simulations see Ref.~\cite{Cea:2004ux}).

% FIGURE 3
\begin{figure}[ht]
\includegraphics[width=0.85\textwidth,clip]{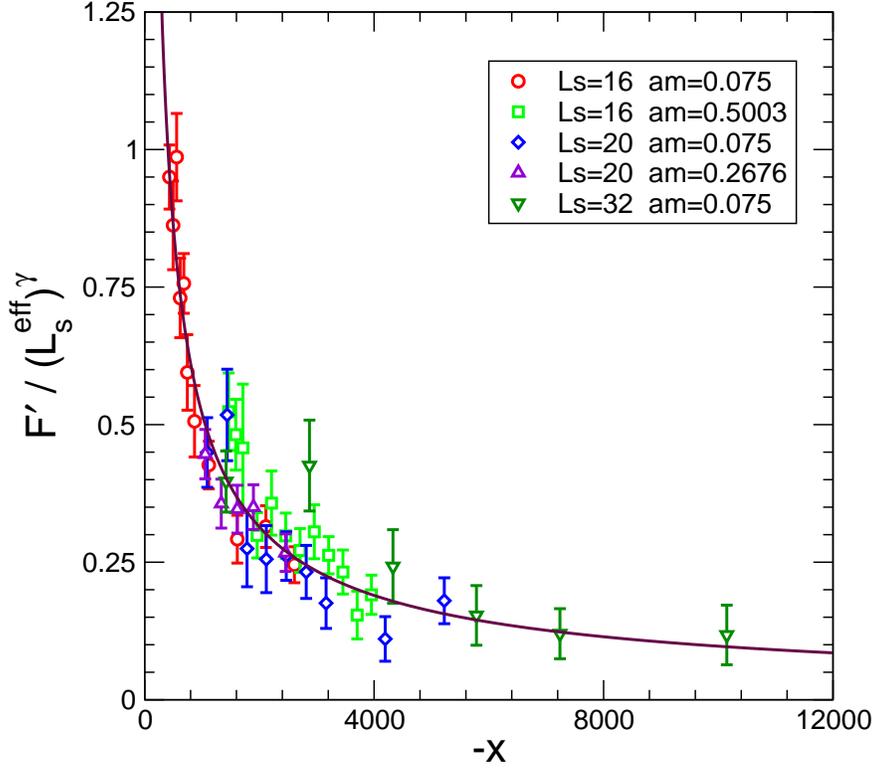}
\caption{
$F^{\prime}(\beta,L_s^{\text{eff}},m_q)$ rescaled by the factor $(L_s^{\text{eff}})^\gamma$.
The values of $L_s^{\text{eff}}$ and $m_q=am$ are displayed in the legend.
}
\label{Fig5}
\end{figure}

We  perform a finite size scaling analysis using the following scaling law (suggested by
Eq.~(\ref{scalinglaw})
\begin{equation}
\label{fssfermions}
F^{\prime}(\beta,L_s^{\text{eff}}) =
\frac{a_1 (L_s^{\text{eff}})^{\gamma}}{\left|  (L_s^{\text{eff}})^{1/\nu} (\beta - \beta_c(m_q)) - d_1 \right|^\alpha} \,,
\end{equation}
where the critical coupling $\beta_c(m_q)$ depends on the quark mass $m_q$.
The dependence of the critical coupling $\beta_c(m_q)$ on the quark mass
is determined by the chiral critical point~\cite{Karsch:1994hm,Engels:2001bq,Karsch:2000kv}.
In the thermodynamical limit, by known universality arguments the critical couplings will scale like
\begin{equation}
\label{betacmq}
\beta_c(m_q) = \beta_c(m_q=0) + c m_q^{1/\beta\delta} \,,
\end{equation}
where $1/\beta\delta$ is a combination of critical exponents which for the case of 2-flavors QCD are
expected to be those of the three-dimensional O(4) symmetric spin models
$1/\beta \delta \simeq 0.5415$.
Inserting Eq.~(\ref{betacmq}) into Eq.~(\ref{fssfermions})  we are lead to the following scaling law
\begin{equation}
\label{scalingfermions}
F^{\prime}(\beta,L_s^{\text{eff}},m_q) =
\frac{a_1 (L_s^{\text{eff}})^{\gamma}}{\left|  (L_s^{\text{eff}})^{1/\nu} (\beta - \beta_c(0) -c m_q^{\eta}) - d_1 \right|^\alpha} \,,
\end{equation}
where again $\gamma = \alpha/\nu$ assures a sensible thermodynamical limit.
Note that, to take care of finite volume effects, the exponent $\eta$ is expected to be
$\eta = \nu_c/\nu$ with $\nu_c = \nu^{\prime}/\beta \delta$, where $\nu^{\prime}$, $\beta$, and  $\delta$ are the chiral critical exponents.
Indeed, Eqs.~(\ref{scalingfermions}) assures that in the scaling region
\begin{equation}
\label{phifermions}
\frac{F^{\prime}(\beta,L_s^{\text{eff}},m_q)}{(L_s^{\text{eff}})^{\gamma}} = \Phi((L_s^{\text{eff}})^{1/\nu}(\beta - \beta_c(0)), (L_s^{\text{eff}})^{1/\nu_c} m_q) \,.
\end{equation}
In our case the relevant chiral critical exponents are those of the three-dimensional O(4) symmetric
spin models where~\cite{Engels:2001bq}
$\nu^{\prime} = 0.7423 \,, \quad \nu_c = 0.4019$.
In Table~\ref{Table3} we report the results obtained by fitting
Eq.~(\ref{scalingfermions}) to all our lattice data.
\begin{table}
\begin{tabularx}{1.0\textwidth}{|XXXXXXXX|}
\hline
\hline
\multicolumn{8}{|c|}{spatial ``fixed boundary conditions'' } \\ \hline
$a_1$   & $\gamma$  &  $\beta_c(0)$     &   $c$       &  $\eta$    &  $\nu$     & $d_1$    & $\alpha$   \\ \hline
$79.4$    & $2.00$    &  $4.9933$       &   $0.54$    &  $1.10$    &  $0.31$    & $0.6$    & $0.728$    \\
$\pm76.6$ & $\pm0.47$ &  $\pm0.0138$    &   $\pm0.11$ &  $\pm0.19$ &  $\pm0.03$ & constant & $\pm0.078$  \\ \hline \hline
\end{tabularx}
\caption{The values of the parameters obtained by fitting Eq.~(\ref{scalingfermions}) to the
data for the derivative of the monopole free energy in two-flavors full QCD
on lattices with spatial volumes $16^3$, $24^3$, and $32^3$ and $L_t=4$.}
\label{Table3}
\end{table}
\\
From Table~\ref{Table3} we can see that $\alpha/\nu=2.35\pm0.34$
consistent with $\gamma=2.00\pm0.47$. Concerning the parameter
$\eta$ we find that it is poorly determined by our data. If we
constrain $\eta$ in our fit we get $\eta = 1.10\pm0.19$ which,
together with $\nu=0.31\pm0.03$ leads to $\nu_c=0.34\pm0.07$
consistent with the value reported above. However if we release the
constraint on $\eta$ our data can also be fitted with smaller values
for $\eta$ without altering significantly the other parameters.
Moreover by confronting the exponent $\alpha$ in Table~\ref{Table3}
with the corresponding value for the SU(3) pure gauge in
Table~\ref{Table2} we conclude that our data for full QCD with two
dynamical flavors are compatible with a first order phase transition
($\nu=0.31\pm0.03$) but this is weaker than in the quenched case.
Our results are in agreement with the indications for a first order
phase transition in full QCD with 2 dynamical flavours (in the same
range of quark masses) obtained
elsewhere~\cite{Carmona:2002yg,Carmona:2002ty,Carmona:2003xs,D'Elia:2005bv,D'Elia:2005ta}.

%%%%%%%%%%%%%%%%%%%%%%%%%%%%%%%%%%%%%%%%%%%%%%%%%%%%%%%%%%%%%%%%%%%%%%%%%%%%%%%%%%%%%%%%%%%%%%%%%%%%%%%%%%%%%%%%%%%%%%%%%%%%%%%%%%%%%
\section{Abelian Chromomagnetic Field}
\label{abchromomagfield}
%%%%%%%%%%%%%%%%%%%%%%%%%%%%%%%%%%%%%%%%%%%%%%%%%%%%%%%%%%%%%%%%%%%%%%%%%%%%%%%%%%%%%%%%%%%%%%%%%%%%%%%%%%%%%%%%%%%%%%%%%%%%%%%%%%%%%

We feel that it could be useful to study vacuum dynamics by using different
perspectives in order to shed light on the basic mechanism of color
confinement.
Indeed,
as recently observed~\cite{'tHooft:2004th} in connection with dual
superconductivity picture, even if magnetic monopoles do condense in
the confinement mode, the actual mechanism of confinement could
depend on additional dynamical forces.

We found~\cite{Cea:1999gn} that a weak constant
abelian chromomagnetic field at zero temperature is completely
screened in the continuum limit, while  at finite
temperature~\cite{Cea:2002wx} our numerical results
indicate  that the applied field is restored by increasing the
temperature. These results strongly suggested that the confinement
dynamics could be intimately related to abelian chromomagnetic gauge
configurations.

We report here numerical results showing that for non abelian gauge theories
the deconfinement temperature depends on the strength of an applied external
constant abelian chromomagnetic field~\cite{Cea:2002wx}.
This is at variance of abelian magnetic monopoles where the abelian monopole background fields
do not modify the deconfinement temperature~\cite{Cea:2004ux}.

%%%%%%%%%%%%%%%%%%%%%%%%%%%%%%%%%%%%%%%%%%%%%%%%%%%%%%%%%%%%%%%%%%%%%%%%%%%%%%%%%%%%%%%%%%%%%%
\subsection{The color vacuum Meissner effect}
%%%%%%%%%%%%%%%%%%%%%%%%%%%%%%%%%%%%%%%%%%%%%%%%%%%%%%%%%%%%%%%%%%%%%%%%%%%%%%%%%%%%%%%%%%%%%%
Let us now define a static constant abelian chromomagnetic field on the lattice.
We consider the SU(3) case in (3+1) dimensions, a thorough discussion
of the SU(2) and U(1) case can be found elsewhere~\cite{Cea:2005td}.

In the continuum the gauge potential giving rise to a static constant abelian chromomagnetic field
directed along spatial direction $\hat{3}$ and direction $\tilde{a}$ in the color space
is given by
\be
\label{su3pot}
\vec{A}^{\text{ext}}_a(\vec{x}) =
\vec{A}^{\text{ext}}(\vec{x}) \delta_{a,\tilde{a}} \,, \quad
A^{\text{ext}}_k(\vec{x}) =  \delta_{k,2} x_1 H \,.
\ee
The constrained lattice links (see Eq.~(\ref{coldwall})) are obtained accordingly to
the continuum gauge potential Eq.~(\ref{su3pot}).
Since our lattice has the topology of a torus,
the magnetic field turns out to be quantized
\be
\label{quant} a^2 \frac{g H}{2} = \frac{2 \pi}{L_1}
n_{\text{ext}} \,, \qquad  n_{\text{ext}}\,\,\,{\text{integer}}\,.
\ee
In the case of a constant background field
the relevant
quantity is the density $f[\vec{A}^{\text{ext}}]$ of
free energy.
We evaluate by numerical simulations the
derivative with respect to the coupling $\beta$
of the free energy density $f[\vec{A}^{\text{ext}}]$ at fixed
external field strength $gH$.

We consider here SU(3) pure gauge theory.
As is well known, the pure SU(3) gauge system undergoes a deconfinement
phase transition at a given critical temperature.
The critical coupling $\beta_c$ can be evaluated by looking at the peak of  $f^{\prime}[\vec{A}^{\text{ext}}]$,
the derivative of the free energy density with respect to $\beta$.

\begin{figure}[ht]
\includegraphics[width=0.65\textwidth,clip]{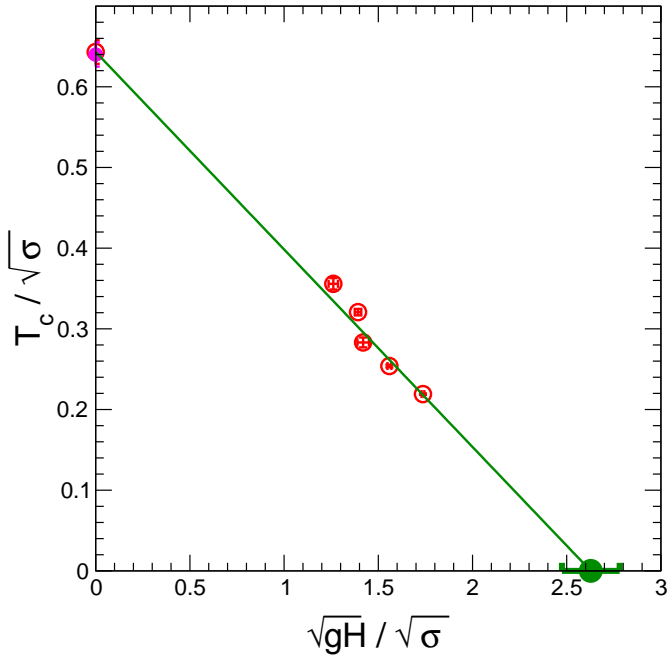}
\caption{SU(3) in (3+1) dimensions. The critical temperature $T_c$ estimated on a $64^3 \times 8$ lattice in units of
the string tension versus the square root of the field strength
$\sqrt{gH}$ in units of the string tension. Solid line is the linear fit eq.~(\ref{tcsqrtsigma}).
In correspondence of zero vertical axis: open circle  is $T_c/\sqrt{\sigma}$ at zero external field;
full circle is the determination of $T_c/\sqrt{\sigma}$ obtained in the literature.
}
\label{Fig4}
\end{figure}

Once the critical coupling $\beta^*(L_t)$ has been determined, the deconfinement
temperature can be given in units of the  string tension
$T_c/\sqrt{\sigma(\beta_c)} = 1/L_t \sqrt{\sigma(\beta_c)}$.
Moreover, using eq.~(\ref{quant}), the field strength  is
$
\sqrt{gH}/\sqrt{\sigma(\beta_c)} = \sqrt{4 \pi n_{\text{ext}}/L_x  \sigma(\beta_c)}
$.
Our data for $T_c/\sqrt{\sigma}$ versus $\sqrt{gH}/\sqrt{\sigma}$ on a
$64^3 \times 8$ lattice are displayed in Fig.~\ref{Fig4}.
It is worth to note that  lattice data can
be reproduced by the linear fit
\be
\label{tcsqrtsigma}
\frac{T_c}{\sqrt{\sigma}} =  \alpha \frac{\sqrt{gH}}{\sqrt{\sigma}} + \frac{T_c(0)}{\sqrt{\sigma}} \,,
\ee
with $T_c(0)/\sqrt{\sigma} = 0.643(15)$ and $\alpha = -0.245(9)$.
The critical field can now be expressed in units of the string tension
and assuming $\sqrt{\sigma}=420$~MeV, the critical field is
$\sqrt{gH_c} = (1.104 \pm 0.063) {\text{GeV}}$
corresponding to $gH_c=6.26(2) \times 10^{19}$~Gauss.

Therefore we may conclude that a critical field exists
such that $ T_c=0$ for $ gH>gH_c$.
This kind of behavior could be interpreted as the colored counterpart of the Meissner effect
in ordinary superconductors, when strong enough magnetic fields
destroy the superconductive BCS vacuum~\cite{Tinkham:1975}.
Then we shall refer to this remarkable result as the reversible color Meissner effect.
It is worthwhile to stress that this effect is not related to the color superconductivity
in cold dense quark matter. Indeed, we believe  that our
reversible color Meissner effect  is deeply rooted in the non-perturbative
color confining nature of the vacuum and could be a window open towards unraveling
the true nature of the confining vacuum.

% FIGURE 4

% FIGURE 5
\begin{figure}[ht]
\includegraphics[width=0.6\textwidth,clip]{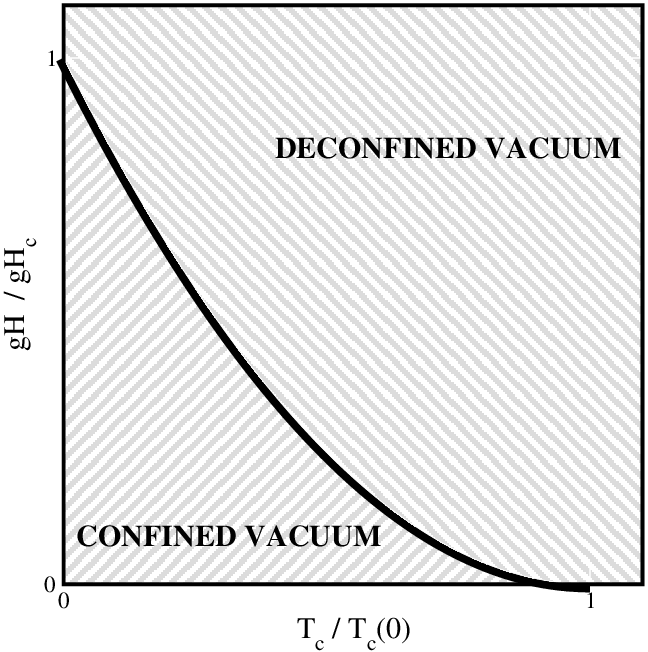}
\caption{Phase diagram of four dimensional SU(2) and SU(3) gauge theories.
}
\label{Fig6}
\end{figure}

%%%%%%%%%%%%%%%%%%%%%%%%%%%%%%%%%%%%%%%%%%%%%%%%%%%%%%%%%%%%%%%%%%%%%%%%%%%%%%%%%%%%%%%%%%%%%%%%%%%%%%%%%%%%%%%%%%%%%%%%%%%%%%%%%%%%%
\section{Conclusions}
\label{conclusions}
%%%%%%%%%%%%%%%%%%%%%%%%%%%%%%%%%%%%%%%%%%%%%%%%%%%%%%%%%%%%%%%%%%%%%%%%%%%%%%%%%%%%%%%%%%%%%%%%%%%%%%%%%%%%%%%%%%%%%%%%%%%%%%%%%%%%%

We investigated the  QCD vacuum using external fields.

We first considered the nature of deconfining phase transition in SU(3) pure gauge theory
and full QCD with two flavors of staggered fermions.
To locate the phase transition
we used the derivative of the monopole free energy with respect to the gauge coupling.
The monopole free energy is defined by means of a gauge invariant
thermal partition functional in presence of the abelian monopole backgroud field.
In the pure gauge case our finite size scaling analysis indicate
a weak first order phase transition.
In the case of 2 flavors full QCD, we performed simulations by varying
spatial lattice sizes and quark masses. We find that deconfinement
transition in full QCD with 2 degenerate dynamical flavors is consistent  with
a weak first order phase transition, contrary to the expectation of a crossover
for not too large quark masses, but in agreement with recent
indications~\cite{Carmona:2002yg,Carmona:2002ty,Carmona:2003xs}.

We then focused our attention to  (3+1) dimensional SU(3) pure gauge theory
in presence of an uniform chromomagnetic field.
We reported numerical evidences in favor of what we named "color vacuum Meissner effect"
that implies the existence of  a critical field
$gH_c$ such that for $gH > gH_c$ the gauge system is the deconfined phase.
Such an effect seems to be generic for non abelian gauge theories~\cite{Cea:2005td} (see Fig.~\ref{Fig6}).
We speculate that the peculiar dependence of the
deconfinement temperature on the strength of the abelian chromomagnetic field
$gH$ could be naturally explained if the vacuum behaved as an ordinary relativistic
color superconductor, namely a condensate of color charged scalar fields whose
mass is proportional to the inverse of the magnetic length. However, the chromomagnetic
condensate cannot be uniform due to gauge invariance of the vacuum, which disorders the
gauge system in such a way that there are not long range correlations.
Consequently  if the vacuum behaved as a non uniform chromomagnetic condensate,
our reversible color Meissner effect
could be easily explained, for strong enough chromomagnetic fields would force long
range color correlations such that the gauge system gets deconfined.
One might thus imagine the confining vacuum in non abelian gauge systems as a disordered
chromomagnetic condensate which confines color charges due both to the presence of
a mass gap and the absence of long range color correlations, as
argued by R.P. Feynman in (2+1) dimensions~\cite{Feynman:1981ss}.

\section*{Acknowledgments}
We acknowledge collaboration with Massimo D'Elia for part of the
results presented in this review.

\section*{References}

%\bibliography{qcd}

\end{document}